\documentclass[aps,prl,twocolumn,10pt,showpacs,letterpaper]{revtex4-1}
\usepackage{graphicx}  
\usepackage{dcolumn}   
\usepackage{amsmath}
\usepackage{amssymb}
\usepackage{mathtools}
\usepackage{epstopdf}
\usepackage{bm}
\usepackage{hyperref}
\usepackage{float}
\usepackage{subfigure}
\usepackage{cleveref}
\usepackage{color}

\usepackage{filecontents,lipsum}

\usepackage{mystyle}
\usepackage{textcase}

\newcommand{\comm}[1]{}
\begin{document}
\title{Nonlinear dispersion relation predicts harmonic generation in wave motion}
\author{Romik Khajehtourian}
\email[Current affiliation: Department of Mechanical and Process Engineering, ETH Z\"urich, Z\"urich, 8092, Switzerland]{}
\author{Mahmoud I. Hussein}
\email[Corresponding author: ]{mih@colorado.edu}
\affiliation{Ann and H.J. Smead Department of Aerospace Engineering Sciences, University of Colorado Boulder, Boulder, Colorado 80309, USA}
\date{\today}

\begin{abstract}

In recent work, we have proposed a theory for the derivation of an exact nonlinear dispersion relation for elastic wave propagation which here we consider for a thin rod (linearly nondispersive) and a thick rod (linearly dispersive). The derived relation has been validated by direct time-domain simulations examining instantaneous dispersion. In this Letter, we present a major extension of the theory revealing that our derived nonlinear dispersion relation provides direct and exact prediction of harmonic generation, thus merging two key tenets of wave propagation. This fundamental unification of nonlinear dispersion and harmonic generation is applicable to any arbitrary wave profile characterized initially by an amplitude and a wavenumber, irrespective of the type and strength of the nonlinearity, and regardless of whether the medium is dispersive or nondispersive in the linear limit. Another direct outcome of the theory is an analytical condition for soliton synthesis based on balance between hardening and softening dispersion.



\end{abstract}


\maketitle

\indent Wave motion lies at the heart of many disciplines in the physical sciences and engineering. For example, natural phenomena involving atomic motion, seismic motion, fluid flow, heat transfer, or propagation of light and sound all involve wave physics at some level \cite{Whitham1974}. While the theory of linear dispersive waves is fairly complete, much has remained to be understood about nonlinear waves and their characterization. For linear systems, it is customary to obtain dispersion relations that relate the frequency $\omega$ and wavenumber $\kappa$ of propagating excitations. A dispersion relation provides valuable information and is often used to characterize a range of physical properties of the medium admitting the wave motion \cite{Nishijima1969}. In nonlinear systems, on the other hand, the notion of a dispersion relation has been treated with caution because superposition does not apply. Yet, the appeal of retaining nonlinear effects in the study of dispersion has motivated several studies in a variety of disciplines, including quantum mechanics \cite{Carretero2008}, solid mechanics \cite{Parker1985,Chakraborty2001,Cobelli2009,Lee2013}, fluid dynamics \cite{Shukla2006, *Onorato2006, *Leoni2014,Herbert2010}, acoustics \cite{Gusev1998}, electromagnetics \cite{Shadrivov2004,*Kourakis2005}, plasma physics \cite{Yoon2003,Huang2009,Ginzburg2010,Hager2012}, geophysics \cite{Fritts2003,*Debnath2007,*Boyd2018}, biophysics \cite{Davydov1973,*Mvogo2018}, among others. Aside from explicit examination of dispersion, classical methods for solving the initial-value problem for a wide class of nonlinear evolution equations have been developed since the late 1960s~\cite{Gardner1967, *Lax1968, *Zakharov1971, *Ablowitz1974}.  \\
\indent In many problems, it is often sufficient to consider the effects of weak nonlinearities; in such cases these effects are augmented over the linear dispersion relation in the form of perturbations \cite{Whitham1965}.~Nonlinear dispersion relations (NDR) for systems exhibiting weak nonlinearities were derived by small-parameter expansions, for example, for discrete chains \cite{Chakraborty2001}, elastic rods \cite{Parker1985}, and plasma \cite{Yoon2003}.~For strong nonlinearities, however, exact derivations of NDR are needed. While rare, a few exact NDR formulations have been produced following what is usually discipline-specific approaches; for example, Sch\"urmann et al.~\cite{Schurmann1988} provided an exact NDR for electromagnetic waves, Huang et al.~\cite{Huang2009} and Ginzburg et al.~\cite{Ginzburg2010} for plasma waves, and Lee et al.~\cite{Lee2013} for Fermi Pasta Ulam waves.\\ 
\indent Yet, a key question remains and that is what does it mean to have a dispersion relation for nonlinear waves, especially when the nonlinearity is strong? The challenge stems from the fact that a nonlinear wave distorts as it travels and appears to ultimately fully lose its original shape and character, and in many instances the final outcome is onset of a form of instability~\cite{Whitham1974}. Inherent to this distortion is an intricate mechanism of harmonic generation, a phenomenon which is in fact widely utilized in laser science \cite{Franken1961}. In the presence of harmonic generation, a Fourier transform of the space-time response reveals a fundamental harmonic as well as a series of weaker but significant higher-order harmonics. The distortion we observe is a manifestation of energy exchange from one harmonic to the other \cite{Hasselmann1962, *Benney1996}. An NDR, as we show, may in principle predict the fundamental-harmonic frequency-wavenumber energy profile that would be generated by plotting a superposition of Fourier spectra of separate wave evolutions excited at the same amplitude but different wavenumbers or frequencies. Aside from dispersion analysis, rigorous theory has been developed for the analysis of the spatial response of multiple harmonics. Early works in this area include papers by Thursten and Shapiro \cite{Thursten1967}, Tiersten and Baumhauer \cite{Tiersten1974}, and Thompson and Tiersten \cite{Thompson1977}. Common techniques applied on weakly nonlinear systems seek spatial solutions for harmonics using perturbation expansions, e.g., see publications on elastic waves by Auld \cite{Auld1973}, Deng \cite{Deng1999}, and de Lima and Hamilton \cite{Lima2003}. This body of theory follow key experimental work on harmonic generation in ultrasonic waves in metals \cite{Breazeale1963, *Hikata1965}. \\
\indent In this Letter, we provide a unification of the concepts of nonlinear dispersion and harmonic generation, and in doing so introduce a new meaning to the notion of an NDR$-$but only when derived following a particular condition. This condition is rooted in a theory introduced in 2013 by Abedinnasab and Hussein \cite{Abedinnasab2013} for the derivation of an exact NDR, which was presented in the context of finite-strain elastic waves in rods and beams. This theory consists primarily of three steps: (1) a traveling phase variable $\xi=\kappa x-\omega t$, where $x$ and $t$ denote position and time, respectively, is substituted into the integrated form of the nonlinear partial differential equation (PDE) governing the wave field; (2) an initial continuous function characterized by amplitude $B$ and wavenumber $\kappa$ is substituted into the PDE; and (3) the condition $\xi=0$ is imposed to eliminate the field variable; this condition restricts the spatial and temporal phases to be equal. Furthermore, secular terms are omitted in the course of the derivation. Upon application to a thin elastic rod exhibiting geometric nonlinearity, this theory generated an exact NDR that was shown to perfectly predict instantaneous dispersion in direct simulations of large-amplitude waves with no restriction on the value of $B$ \cite{Abedinnasab2013}. A rod is considered thin when its thickness is much smaller than the wavelengths of propagating waves.\\
\indent In the present work, we again examine a thin elastic rod, which in the linear limit of $B \rightarrow 0$ is described by a nondispersive $\omegaω-\kappa$ relation, but also consider a thick elastic rod, which in the linear limit exhibits dispersion due to the effect of lateral inertia. \\
\indent \textit{Thick elastic rod}$-$We consider an infinite one-dimensional (1D) rod with polar radius of gyration $r$ and constant material properties. The rod admits longitudinal displacements $u(x,t)$ under uniaxial stress $\sigma(x,t)$. The governing equation of motion (EOM) is obtained by Hamilton's principle $\delta\int\nolimits_{0}^{t}\mathcal{L}\partial t=0$. 
In the absence of external non-conservative forces and moments, the Lagrangian density function $\mathcal{L}={\mathcal{T}}-{\mathcal{U}}$ encapsulates the dynamics of the system via the kinetic and strain energy densities ${\mathcal{T}}=\rho \lbrack(\partial_{t}{{u}})^2+\nu^2 r^2(\partial^2_{tx}{{u}})^2\rbrack/2$ and  ${\mathcal{U}}=\sigma\epsilon/2$, respectively. Here, $\rho$ and $\nu$ denote the mass density and Poisson's ratio, and the stress follows Hook's law $\sigma= E\epsilon$, where $E$ is the elastic modulus. The finite strain $\epsilon$ may in principle take the form of any of the Seth-Hills family of strain measures \cite{Seth1962, *Hill1968}. Here we consider, separately, the  Green-Lagrange strain (GLS) and Hencky strain (HS) measures which are defined as $\epsilon=\partial_{x}{u}+(\partial_{x}{u})^2/{2}$ and $\epsilon=\ln(1+\partial_{x}{u})$, respectively.~Applying the functional form $\mathcal{L}$ to the Euler-Lagrange equation $\partial_{x} (\partial_{\partial_{x}{u}}\mathcal{L})+\partial_{t} (\partial_{\partial_{t}{u}}\mathcal{L})=0$, followed by defining $\bar{u}=\partial_x{u}$, yields the EOM for the displacement gradient. This second-order partial differential equation takes the general form:
\begin{equation}\label{eq2:PDE}
\partial_{tt}\bar{u}-\partial_{xx}(\alpha \bar{u}+\beta \mathcal{N}(\bar{u})+\gamma \partial_{tt}\bar{u})=0,
\end{equation}
where $\alpha=\beta=c^2$ and $\mathcal{N}(\bar{u})=3{\bar{u}}^2/2+{\bar{u}}^3/2$ for the GLS measure, and $\alpha=0$, $\beta=c^2$, and $\mathcal{N}(\bar{u})=\ln(1+\bar{u})/(1+\bar{u})$ for the HS measure. 
For both cases, the quasistatic speed of sound is given by $c=\sqrt{E/\rho}$, and for compactness we have introduced the parameter  $\gamma=r^2 \nu^2$.\\
\indent For the limiting configuration of a thin rod, the lateral inertia is omitted by setting $r=0$. In this medium, a traveling wave profile with finite-amplitude $B$ will experience in the course of its evolution forward self-steepening in the case of Green-Lagrange nonlinearity and backward self-steepening in the case of Hencky nonlinearity.~Eventually each wave experiences instability at time $\tau_{\rm B}$.~Both scenarios are demonstrated in Fig.~\ref{Fig1a}, and an analysis of this steepening effect and its path to instability is given in Appendix A.\\ 
\indent \textit{General nonlinear dispersion relation}$-$Following the theory presented in Ref.~\cite{Abedinnasab2013}, we derive the NDR for a rod but now also account for the linear effect of lateral inertia~\footnote{For simplicity, we consider the effect of the lateral inertia only on the longitudinal displacement; however the theoretical framework is fully valid in the absence of this modeling simplification.}, and consider the case of HS in addition to the case of GLS. As summarized above, a change of variables $\xi=\kappa x-\omega t$ is first introduced. This transforms Eq.~(\ref{eq2:PDE}) to
\begin{equation}\label{eq:ODE}
\omega^2\partial_{\xi\xi}\bar{u}-\kappa^2\partial_{\xi\xi}(\alpha \bar{u}+\beta \mathcal{N}(\bar{u})+\omega^2\gamma \partial_{\xi\xi}\bar{u})=0,
\end{equation}
which upon integration twice yields
\begin{equation}\label{eq:Algebraic}
(\kappa^2\alpha-\omega^2)\bar{u}+\kappa^2\beta \mathcal{N}(\bar{u})=0.
\end{equation}
The integration constants leading to Eq.~(\ref{eq:Algebraic}) have been set equal to zero, which is consistent with a bounded traveling wave solution for an initially bounded displacement field. Next we assume a displacement field characterized by $B$ and $\kappa$ and that satisfies the condition ${\bar{u}}(0)=B\kappa$ at $\xi=0$. The simplest choice that meets these criteria is $u(\xi)=B \mathrm{sin}(\xi)$, which gives $\bar{u}(\xi)=\kappa B \mathrm{cos}({\xi})$. Upon application of this solution form and condition to Eq.~(\ref{eq:Algebraic}), we obtain the exact NDR respectively for the GLS and HS measures as:
\begin{eqnarray}\label{eq:w-GL}
		\omega^{\mathrm{GLS}}=c\kappa\sqrt{({2+3B\kappa+B^2 \kappa^2})/(2+2  \gamma \kappa^2)},
\end{eqnarray}
and
\begin{eqnarray}\label{eq:w-H}
	\omega^{\mathrm{HS}}=c\kappa\sqrt{{\ln (1+B\kappa)}/({B\kappa(1+B\kappa)(1+\gamma \kappa^2)})}.
\end{eqnarray}
These reduce to a linear dispersive form in the limit of $B\rightarrow 0$ and a linear nondispersive form in the limit $B\rightarrow 0, \gamma\rightarrow 0$.~Figure~\ref{Fig1c} presents plots of the general NDR, as defined in Eqs.~(\ref{eq:w-GL}) and ~(\ref{eq:w-H}), for a thin rod and $B=0.1$. It is noteworthy that nonlinearity by itself causes wave dispersion~\cite{Abedinnasab2013}. \\
    \begin{figure}[!t]
    \centering
     \subfigure{\includegraphics[scale=1]{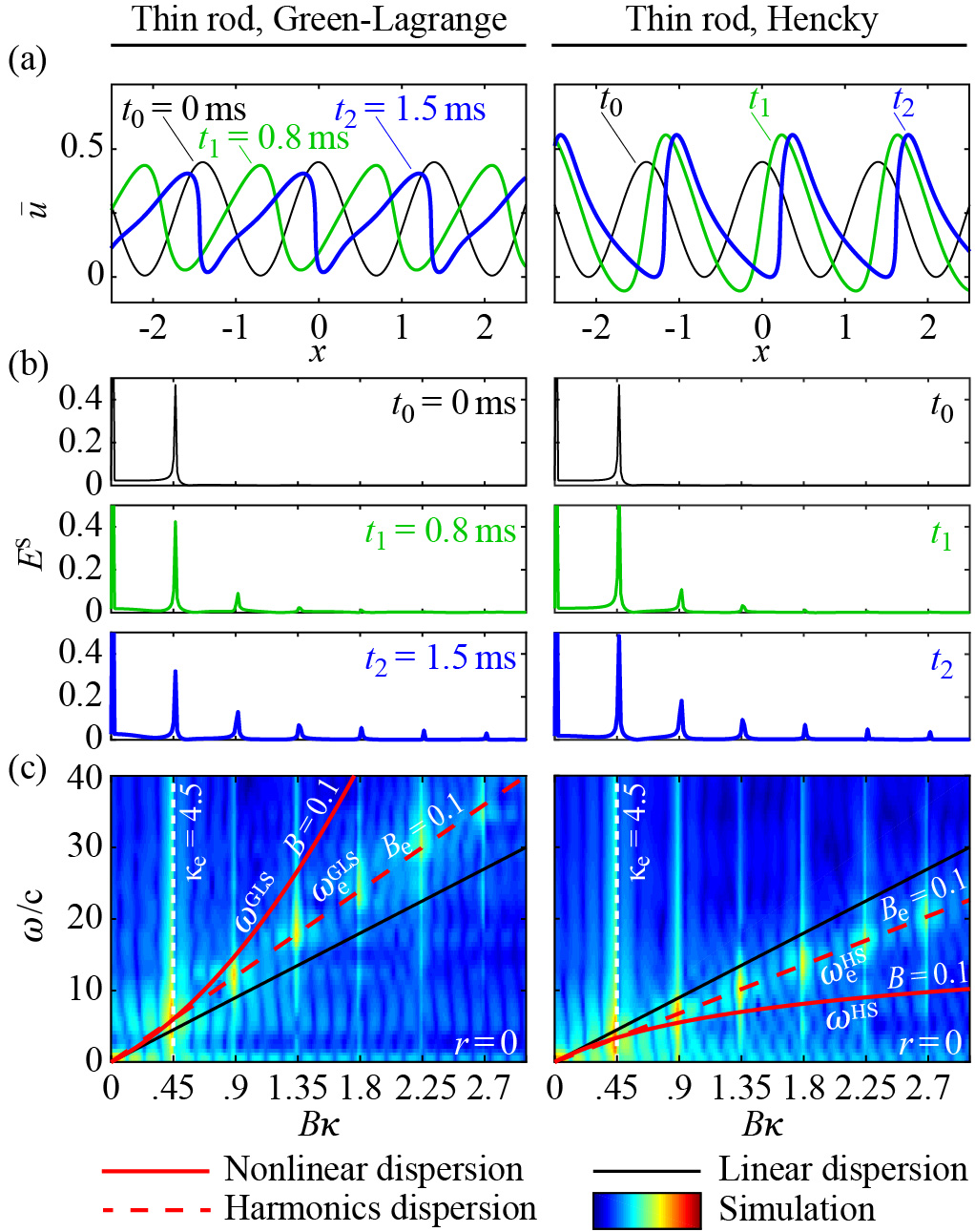}\label{Fig1a}}
     \subfigure{\label{Fig1b}}
     \subfigure{\label{Fig1c}}
    		\caption{(color).~Cosine wave experiencing distortion and harmonic generation. (a) Spatial profiles captured at three different times. (b) Wavenumber and (c) frequency-wavenumber spectrum of harmonics demonstrating perfect prediction by the harmonics dispersion relations of Eqs.~(\ref{eq:GL_ke}) and~(\ref{eq:H_ke}), respectively.
    		\label{Fig1}
    		}
    \end{figure}
      \begin{figure*}
       \centering
       \subfigure{\includegraphics[scale=1]{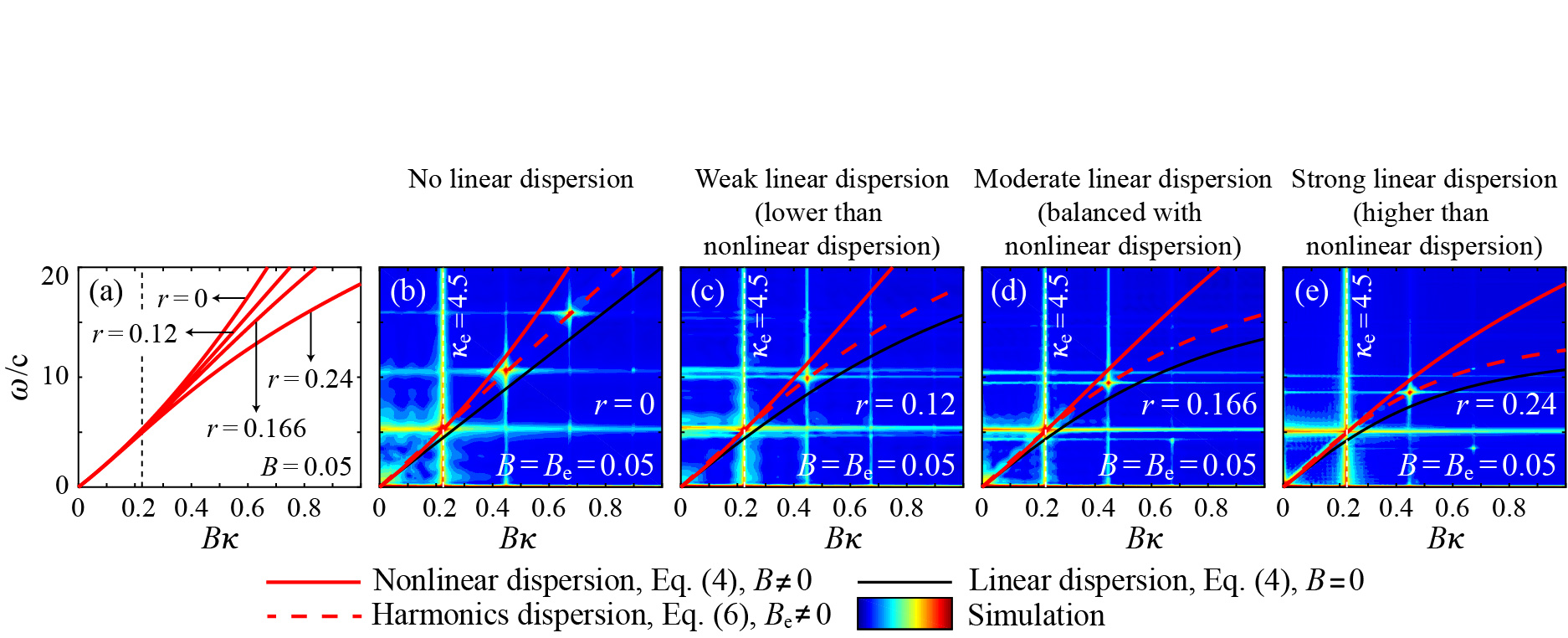}\label{Fig2a}}
       \subfigure{\label{Fig2b}}
       \subfigure{\label{Fig2c}}
       \subfigure{\label{Fig2d}}
       \subfigure{\label{Fig2e}}	
       \caption{(color).~Harmonics dispersion relation predicts harmonics generated by direct simulations for both linearly nondispersive ($r=0$) and linearly dispersive ($r\neq0$) rods. (a) Effect of lateral inertia on the general NDR. (b-e) Frequency-wavenumber spectrum of harmonics with general NDR and harmonics dispersion relation for each case overlaid. Balance of linear and nonlinear dispersion is demonstrated in (d). All results are based on the GLS measure. \label{Fig2}}
      \end{figure*}
\indent \textit{Harmonics dispersion relation}$-$We now uncover that the general NDR derived above inherently encompasses information on the harmonic generation mechanism associated with nonlinear waves characterized initially by an amplitude $B=B_{\rm e}$ and a wavenumber $\kappa=\kappa_{\rm e}$. Starting with Eq.~(\ref{eq:Algebraic}), we impose the condition ${\bar{u}}(0)=B_{\rm e}\kappa_{\rm e}$, which yields the following exact \it harmonics dispersion relation\rm~for the GLS and HS cases, respectively:
\begin{equation}\label{eq:GL_ke}
\omega^{\mathrm{GLS}}_{\mathrm{e}}=c\kappa\sqrt{({2+3B_{\mathrm{e}}\kappa_e+B_{\mathrm{e}}^2 {\kappa_e}^2})/{(2+2  \gamma \kappa^2)}},
\end{equation}
and 
\begin{equation}\label{eq:H_ke}
\omega^{\mathrm{HS}}_{\mathrm{e}}=c\kappa\sqrt{{\ln (1+B_{\mathrm{e}}\kappa_e)}/{(B_{\mathrm{e}}\kappa_e(1+B_{\mathrm{e}}\kappa_e)(1+\gamma \kappa^2))}}.
\end{equation}
Each of these relations predicts the exact frequency-wavenumber curve on which all the harmonics will lie following a Fourier transform of a spatially evolving nonlinear pulse of any arbitrary form provided that the pulse initially satisfies the condition ${\bar{u}}(0)=B_{\rm e}\kappa_{\rm e}$. If the pulse is not balanced, e.g., is experiencing self-steepening, the prediction will be valid up to the point of onset of instability.~Figure~\ref{Fig1c} presents plots of the harmonics dispersion relation, as defined in Eqs.~(\ref{eq:GL_ke}) and ~(\ref{eq:H_ke}), for a thin rod for $B_{\rm e}=0.1$ and $\kappa_{\rm e}=4.5$. The notion of a harmonics dispersion relation represents a new paradigm in nonlinear wave science.\\
\indent \textit{Validation by direct simulations}$-$Here we seek a numerical solution of Eq. (\ref{eq2:PDE}) to validate our assertion that each of Eqs.~(\ref{eq:GL_ke}) and~(\ref{eq:H_ke}) (for the GLS and HS measures, respectively) represent a dispersion relation for harmonic generation.~We use a spectral method in conjunction with an efficient explicit time-stepping method to obtain the response as a function of position and time. Afterwards, a discrete Fourier transform is performed on the simulated space-time field to reveal the spectrum of the emerging harmonics and compare their distribution in the frequency-wavenumber domain with the analytically derived harmonics dispersion relation (see Appendix B for details on the numerical approach).\\
\indent We consider a periodic domain and prescribe initially an ``excitation" harmonic wave, or wave packet, that is characterized by an amplitude $B_{\rm e}$ (arbitrarily normalized) and a wavenumber $\kappa_{\rm e}$.
In principle, any arbitrary but well-defined and smooth wave function $\bar{u}(x,t)$ satisfying $\bar{u}(\xi=0)= B_{\rm e}\kappa_e$, the condition used to derive the general NDR, may be used in these simulations. The space-time domain is defined by $-x^*<x\leq x^*$ (large enough to avoid any reflections from the boundaries) with a grid spacing of $h=1$~mm, and $0\leq t\leq \tau_\mathrm{B}$ with a constant time step of $\Delta t=1~\mu$s. The material properties considered are for aluminum: $\rho=2700$~kg/m$^3$, $E=70$~GPa, and $\nu=0.33$~\footnote{All reported units are in the SI system.}. \\
\indent First, we examine a simple cosine wave profile as an excitation signal and apply it to the case of the thin rod. This wave profile has the form $\bar{u}(x,t)= B_{\rm e}\kappa_{\rm e}[1+\cos(\kappa_{\rm e}(x-ct))]/2$ and is characterized by $B_{\rm e}=0.1$ and $\kappa_{\rm e}=4.5$. We prescribe this $\bar{u}$ field along the entire computational domain at time $t=0$. Beyond the initial time, the wave is allowed to propagate freely in the simulation with no further prescription of displacement. Three time snapshots of the simulated motion are shown in Fig. \ref{Fig1a}.\\  
\indent Performing Fourier analysis in space on the wave function at the time of excitation ($t_{\rm 0}$) and at two further times ($t_{\rm 1}$ and $t_{\rm 2}$) shows the evolution of the energy spectrum, $E^{\rm S}$. At $t_{\rm 0}$, only a single harmonic exists (which is of the cosine excitation signal). As the wave evolves, the nonlinear effects increasingly cause distortion and generation of higher harmonics, as shown in Fig. \ref{Fig1b}. In Fig. \ref{Fig1c}, we show a representation of the contour of the energy spectrum $E^{\rm ST}$ produced by Fourier analysis in both space and time~\footnote{All displayed numerical contour plots are of the quantity $\mathrm{ln}{ |E^{\rm ST}_{B_{\rm e},\kappa_{\rm e}}|}$ (energy spectrum due to an excitation at $B_{\rm e}$ and $\kappa_{\rm e}$) or $\sum_{\kappa_{\rm e}}[\mathrm{ln}{ |E^{\rm ST}_{B_{\rm e},\kappa_{\rm e}}|}]$ (superposition of energy spectra over several values of $\kappa_{\rm e}$ for a given value of $B_{\rm e}$). The Fourier transformation is always done at a time close to $\tau_\mathrm{B}$. Details on the derivation of $E^{\rm ST}$ are provided in Appendix A.}. The brightened areas in the contour plot represent the harmonics, and these are observed to perfectly coincide with the harmonics dispersion curve derived in Eqs.~(\ref{eq:GL_ke}) (left panel) and (\ref{eq:H_ke}) (right panel), thus providing confirmation of the theory. \\
\begin{figure*}
  \centering
  \subfigure{\includegraphics[scale=1]{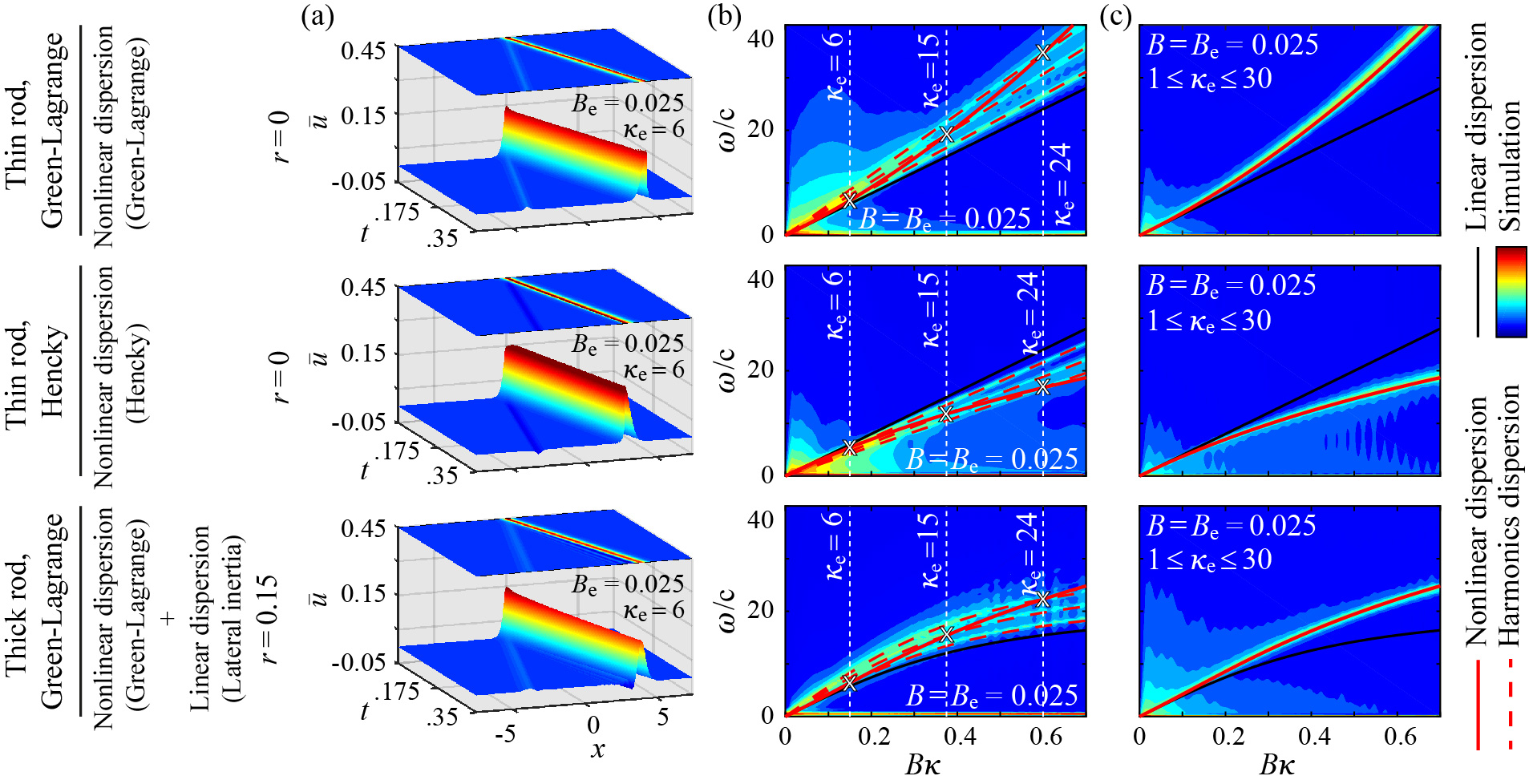}\label{Fig3a}}
  \subfigure{\label{Fig3b}}
  \subfigure{\label{Fig3c}}
  \caption{(color).~Demonstration of the connection between general NDR and harmonics dispersion relation. (a) Space-time numerical simulation of a large-amplitude hyperbolic secant wave profile. (b) Frequency-wavenumber spectra showing distribution of harmonics in evolved field. The intersections of the harmonics dispersion relations (dashed red curves) for three distinct excitation wavenumbers are shown to coincide perfectly with the general NDR for the selected value of wave amplitude. (c) Superposition of harmonics spectra from thirty distinct simulations covering a range of excitation wavenumbers is shown to match perfectly with the general NDR for the selected value of wave amplitude. Different cases are considered in each of the top, middle, and bottom panels.\label{Fig3}}
 \end{figure*}
\indent We also consider a thick rod under the GLS measure and propagate the same excitation but now characterized by $B_{\rm e}=0.05$ and $\kappa_e=4.5$ with various rod thicknesses. Unlike conventional techniques such as the method of characteristics which fall short in the presence of linear dispersion \cite{ablowitz2011nonlinear}, our theory predicts harmonic generation even for systems that are linearly dispersive (e.g., $r\neq0$), as is shown in Fig. \ref{Fig2}. We observe that for all $r\neq0$ cases [Figs. \ref{Fig2}(c-e)], the harmonics dispersion relation exhibits dispersion$-$indicating a nonlinear softening trend for the generated harmonics in line with the dispersive nature stemming from lateral inertia. \\
\indent Next we simulate the nonlinear wave propagation of a localized pulse defined by $\bar{u}(x,t)= B_{\rm e}\kappa_e[\text{sech}(\kappa_e(x-ct))]$.~Figure \ref{Fig3a} shows the results for $B_{\rm e}=0.025$ and $\kappa_e=6$.~The initial wave packet propagates along the positive direction; also observed is a characteristic trailing wave of modest amplitude radiating in the opposite direction~\footnote{The spatial profile of the waves feature the eventual formation of shocks at the leading and trailing edges of the wave packet for the GLS and HS measures, respectively$-$in analogy with the behavior observed in Fig.~\ref{Fig1a}.}.~In Fig. \ref{Fig3b}, we consider excitations at two more $\kappa_e$ values and superpose the Fourier-transformed spectra for all three cases.~It is seen that the harmonics dispersion curves perfectly predict the distribution of the harmonics for each of the excitations, including the linearly dispersive case shown in the bottom panel. It is noteworthy that since the hyperbolic secant function has a rich frequency content, the excited energy spectrum for each of the three cases displayed in Fig.~\ref{Fig3b} contiguously conform to the harmonics dispersion relations plotted in dashed red.  Furthermore, the set of the intersections of each harmonics dispersion curve with its corresponding $\kappa_e$ value exactly follow the path of the general NDR plotted in red. We learn from this perfect matching of intersections that the general NDR curve traces the fundamental harmonic associated with each excitation wavenumber (for a given value of $B_{\rm e}$). This characteristic is confirmed further in Fig.~\ref{Fig3c} by superimposing the energy spectra of thirty separate simulations for distinct initial wave packets sharing the same amplitude but covering the range of excitation wavenumbers $\kappa_e=$1 to 30, with increments of $1$.~The results of Figs. \ref{Fig3b} and \ref{Fig3c} confirm that our unified theory holds for arbitrary excitation profiles such as the hyperbolic secant function considered. \\ 
\begin{figure}[!b]
	\centering
	\subfigure{\includegraphics[scale=1]{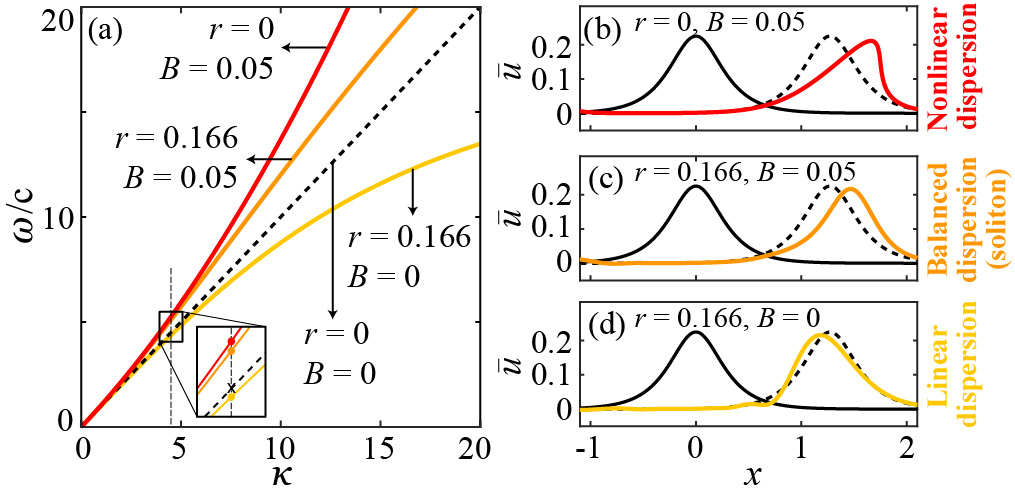}\label{Fig4a}}
	\subfigure{\label{Fig4b}}
	\subfigure{\label{Fig4c}}
	\subfigure{\label{Fig4d}}
	\caption{(color).~Illustration of soliton synthesis in a thick rod by (a) balance of hardening (nonlinear) dispersion with softening (linear) dispersion. The solid black curves represent an initial hyperbolic secant wave; the corresponding colored curves represent the evolved wave for (b) nonlinear dispersion, (c) balanced dispersion (soliton), and (d) linear dispersion. \label{Fig4}}
\end{figure}
\indent \textit{Soliton synthesis}$-$Solitons research traces back to its first observation in a canal by J.S. Russel~\cite{Russel1844} and key early theoretical developments that followed~\cite{Korteweg1895,*Adlam1958}. Other than their solitary spatial profile, a unique aspect of this class of waves is their inherent stability which is commonly attributed to a balance between nonlinear and dispersion effects~\cite{ablowitz2011nonlinear}.~Using the general NDR, we are able to find a condition for balance between hardening dispersion (stemming from the nonlinear kinematics) and softening dispersion (stemming from the linear lateral inertia). This represents a formal approach for the study and engineering of solitons. This approach has been presented recently by the authors in the context of 1D periodic rods~\cite{Hussein2018}.~For a thick rod, we formulate the \it soliton synthesis condition \rm as
\begin{widetext}
	\begin{equation}\label{eq:balance} 
	\begin{aligned}
	r\approx\arg\min\{||\omega_{\mathrm{GLS}}/c-\alpha B\kappa||_{[0,1]}/\max_{[0,1]}(\omega_{\mathrm{GLS}}/c)<1{\%},
	\alpha\in\mathbb{R}\},
	\end{aligned}
	\end{equation} 
\end{widetext}
which gives an optimal value of $r=0.166$ for $B=0.05$. These values generate a linear-nonlinear dispersion balance within 1\% error for the range $0\leq B\kappa\leq 1$, which is the case displayed in  Fig.~\ref{Fig2d} where the NDR appears nearly as a perfect linear curve. Figure ~\ref{Fig4a} decomposes the balancing components and Figs.~\ref{Fig4}(b-d) illustrate the effect in the spatial domain showing, in Fig.~\ref{Fig4c}, the stable propagation of a synthesized soliton. \\
\indent \textit{Conclusions}$-$We have provided a unified theory of nonlinear waves consisting of a general NDR and a harmonics dispersion relation, where the former encompasses the latter. A general NDR defines for a given amplitude (1) the instantaneous dispersion of a nonlinear wave and (2) the frequency-wavenumber spectrum of the fundamental harmonic in a superposition of evolved nonlinear wave fields spanning a range of excitation wavenumbers.~Prescription of a condition for soliton synthesis is a natural outcome of the general NDR, as demonstrated in Fig.~\ref{Fig2}(d) and Fig.~\ref{Fig4}.~A harmonics dispersion relation defines the frequency-wavenumber spectrum of the generated harmonics in an evolved nonlinear wave field for a given excitation amplitude and wavenumber.~There is no limitation by the type and strength of the nonlinearity, nor by the shape of the wave profile.~There is also no restriction on the presence of linear dispersion.~The theory is in principle applicable to other types of waves beyond elastic waves.

\section{A\lowercase{cknowledgments}}
\begin{acknowledgments}
\indent~This work was partially supported by the National Science Foundation CAREER Grant No. 1254937. The authors express their utmost gratitude to Professor M.J. Ablowitz for sharing his experience and insights and for helping them see where their work sits within the full scope of nonlinear wave science. The authors are also thankful to Professors M.A. Hoefer and C.A. Felippa for fruitful discussions.
\end{acknowledgments}

\appendix

\section{\label{sec:AppA} Appendix A: Wave steepening and stability analysis}
\renewcommand{\theequation}{A\arabic{equation}} 
\renewcommand{\thefigure}{A\arabic{figure}} 
\renewcommand{\thesection}{A\arabic{section}}   
\setcounter{figure}{0}
\setcounter{equation}{0} 
\setcounter{section}{0} 
\indent To investigate the nonlinear wave distortion and breaking phenomena, we consider the limiting configuration of a thin rod where the lateral inertia is ignored by setting $r=0$. In this medium, a traveling wave profile with finite-amplitude $B$ will experience in the course of its evolution forward self-steepening in the case of Green-Lagrange nonlinearity and backward self-steepening in the case of Hencky nonlinearity.~In forward steepening, the leading edge has a steeper slope than the trailing edge; and vice versa in backward steepening.~Thus Green-Lagrange-induced steepening takes place in the direction of propagation and represents a \it dispersion hardening\rm~effect \cite{Hussein2018} that eventually leads to the formation of a shock $\partial_x \bar{u}=-\infty$.~In contrast, Hencky-induced steepening causes a tilt opposite to the direction of propagation and represents a \it dispersion softening\rm~effect~\cite{Remillieux2016} that eventually leads to the formation of a shock $\partial_x \bar{u}=\infty$. Both scenarios are demonstrated in Fig.~\ref{FIGSM1}. Consider $m(x,t)<0$ and $M(x,t)>0$ to be the minimum and maximum value of $\partial_x \bar{u}$ as a function of time. There exist a finite bifurcation time $\tau_\mathrm{B}$ when at least one point of the wave profile slope becomes vertical, $m\rightarrow -\infty$ and  $M\rightarrow \infty$, and a shock forms at the leading edge in the GLS case and at the trailing edge in the HS case~\cite{Dai2000}.~These effects are quantified by characteristic lines, as demonstrated in the inset of Fig.~\ref{FIGSM1}.\\
\indent The stability of this nonlinear thin rod can be locally evaluated using eigenvalues of Eq.~(\ref{eq:ODESYS}) which is the equivalent first-order system of Eq.~(2) ignoring the effects of lateral inertia,
\begin{equation}\label{eq:ODESYS}
\begin{aligned}
\partial_{\xi}\bar{u}=\bar{v},\\
\partial_{\xi}\bar{v}=\frac{\kappa^2}{\omega^2}\partial_{\xi\xi}(\alpha \bar{u}+\beta \mathcal{N}(\bar{u})).
\end{aligned}
\end{equation}
 \begin{figure}[!t]
 	\centering
 	\includegraphics[scale=1]{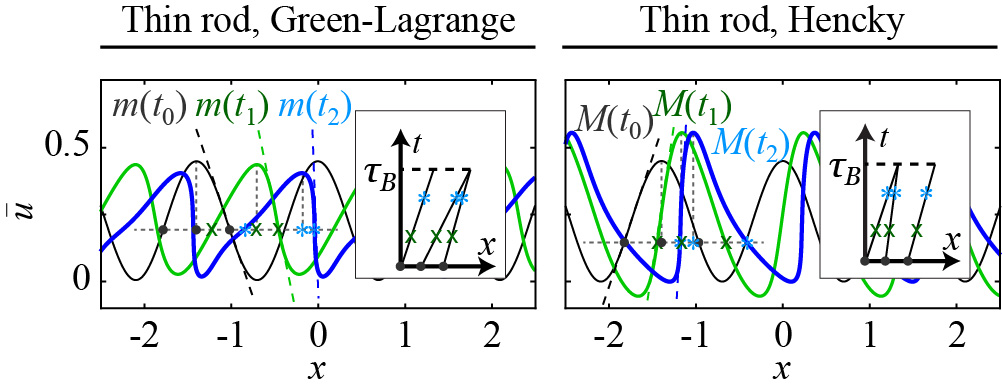}
 	\caption{(color).~Cosine wave experiencing distortion and harmonic generation; spatial profiles captured at $t_0=0$ ms, $t_1=0.8$ ms, and $t_2=1.5$ ms. Inset shows the corresponding characteristic lines.
 		\label{FIGSM1}
 	}
 \end{figure}
By analyzing this system, we find two distinct real eigenvalues, one negative and one positive, which collide into each other on $\bar{v}=0$. The positive eigenvalue indicates that the system is unstable and the solutions are in the form of breaking waves.\\
\indent The position and time of bifurcation may be determined by solving $\partial_{\bar{u}} x(\bar{u})=0$ (equivalent to $|\partial_x \bar{u}(x)|=\infty$) for a known solution, and setting $\partial_{\bar{u}\bar{u}} x(\bar{u})=0$ as a necessary condition to ensure the uniqueness of $\bar{u}(x)$.\\

\section{Appendix B: Computational approach description}
\renewcommand{\theequation}{B\arabic{equation}}
\renewcommand{\thefigure}{B\arabic{figure}}  
\renewcommand{\thesection}{B\arabic{section}}  
\setcounter{figure}{0}
\setcounter{equation}{0} 
\setcounter{section}{0} 
\begin{figure}[!b]
	\centering
	\subfigure{\includegraphics[scale=1]{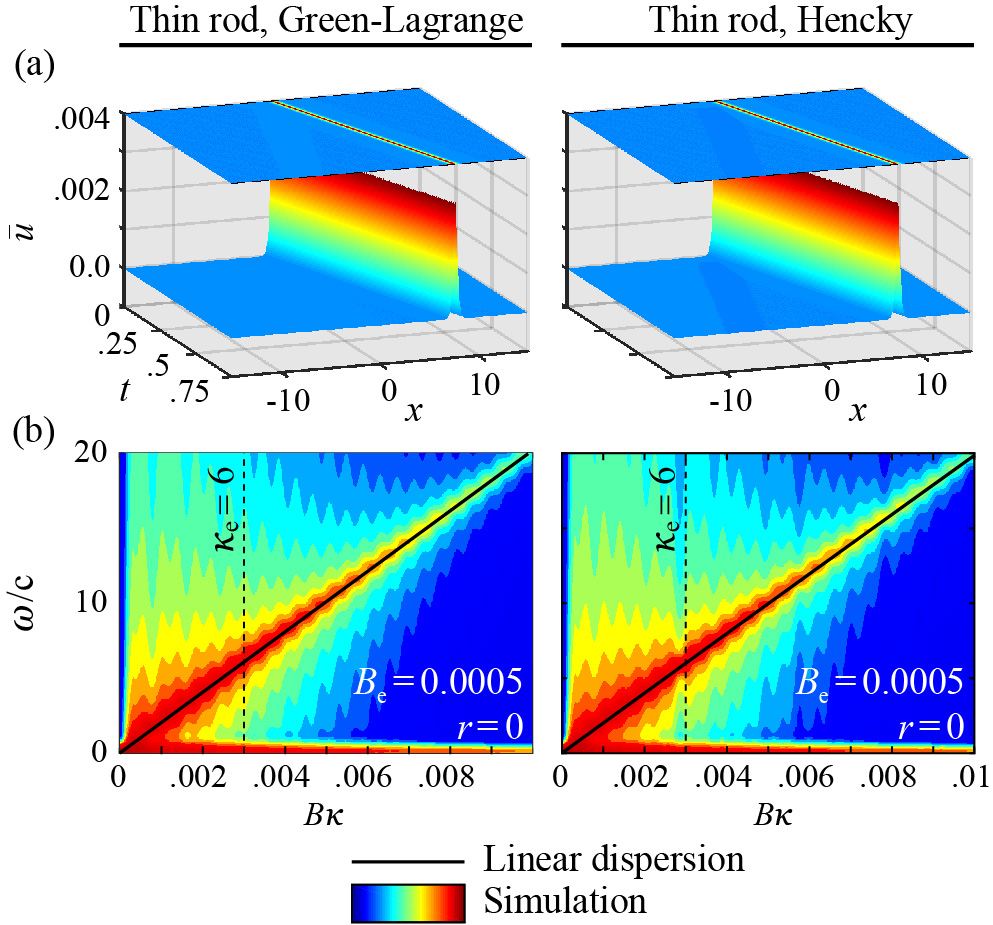}\label{FigSM2a}}
	\subfigure{\label{FigSM2b}}
	\caption{(color). Verification of computational approach on a linear nondispersive rod (GLS measure in left column and HS measure in right column).~(a) Infinitesimal strain space-time solution for $B_{\rm e}=0.0005$ and $\kappa_{\rm e}=6$. (b) A representation of the energy spectrum $E^{\rm ST}$ obtained by Fourier transformation. Corresponding dispersion curves from Eqs.~(4) and~(5) are overlaid as solid lines. Time and space units are [ms] and [m], respectively.\label{FigSM2}}
\end{figure}
\indent We simulate the propagation of Eq.~(1) using a spectral method for the spatial variable in conjunction with an efficient explicit time-stepping method. The nonlinear PDEs are discretized with the discrete Fourier transform (DFT) in space and marched in time using a numerical integration scheme.~We consider $\bar{u}_j$ as a discrete function on an $N$-point spatial grid $x_j$, $j=1,\dots,{N}$. The DFT is defined by
$\widehat{u}_{k}=h\sum\nolimits_{j}\mathrm{e}^{-\mathrm{i}kx_j}\bar{u}_j, $ for $k=-{N}/{2}+1,...,{N}/{2},$ and the inverse discrete Fourier transform (IDFT) by ${\bar{u}}_{j}=\frac{1}{2\pi}\sum\nolimits_{k}\mathrm{e}^{\mathrm{i}kx_j}\widehat{u}_k, $ for each point. Here, $h=2\pi/N$, $x_j=jh$, $h$ is the spacing of the grid points, and $k$ is the Fourier wavenumbers.
We apply $\partial_t \bar{u}=\bar{v}$ followed by the DFT on Eq.~(1) to form the corresponding first order system
\begin{equation}\label{eq:ODESYSDFT}
 \partial_t\left[\begin{array}{cc} \widehat{u}\\\widehat{v} \end{array}\right]
      =\begin{bmatrix}        0 & 0\\ -\frac{\alpha k^2}{1+\gamma k^2}& 0    \end{bmatrix}\left[\begin{array}{cc} \widehat{u}\\\widehat{v} \end{array}\right]+\begin{bmatrix}        \widehat{v} \\-\frac{\beta k^2}{1+\gamma k^2} \mathcal{F}(\mathcal{N})    \end{bmatrix},
\end{equation}
where $\mathcal{F}(.)$ denotes the Fourier transform of the considered function.~Differentiating the transformation $ \left[{\widetilde{u}},{\widetilde{v}}\right]^\textrm{T}=\bm{\Gamma}\left[\widehat{u},\widehat{v}\right]^\textrm{T}$ with respect to time, with
$\bm{\Gamma}=[I,0;{\alpha k^2}\Delta t , I]$ being the integral factor of Eq.~(\ref{eq:ODESYSDFT}), followed by the substitution of the $\partial_t{\widehat{u}}$ and $\partial_t{\widehat{v}}$ values from Eq.~(\ref{eq:ODESYSDFT}) and $\widehat{u}$ and $\widehat{v}$ from the inverse transformation $\left[\widehat{u},\widehat{v}\right]^\textrm{T}=\bm{\Gamma}^{-1}\left[{\widetilde{u}},{\widetilde{v}}\right]^\textrm{T}$, produces the following numerically integrable system (returning to continuous notation for convenience):
\begin{equation}\label{eq:ODESYSInt}
  \begin{aligned}
\partial_t\widetilde{u}=-\frac{\alpha k^2 \Delta t}{1+\gamma k^2}\widetilde{u}+\widetilde{v},\\
\partial_t\widetilde{v}=\frac{\alpha k^2 \Delta t}{1+\gamma k^2}(-\frac{\alpha k^2 \Delta t}{1+\gamma k^2}\widetilde{u}+\widetilde{v})-\frac{\beta k^2}{1+\gamma k^2}\mathcal{F}(\mathcal{N}).
  \end{aligned}
\end{equation}
\indent We use the fourth-order explicit Runge-Kutta time-stepping scheme to integrate Eq. (\ref{eq:ODESYSInt}).~Then the inverse transformation is applied followed by IDFT on $ \left[{\widetilde{u}},{\widetilde{v}}\right]^\textrm{T}$ to obtain $\bar{u}(x, t)$.~The direction of wave propagation in the simulation is dictated by the initial velocity condition we prescribe.~Now that we have the space-time solution, we apply Fourier analysis to the spatio-temporal wave-field discrete data ${\bar{u}}_{p,q}, p=0, 1,..., N-1,  q=0, 1,..., T-1$ by $E^{\rm ST}_{l, n}=\frac{1}{NT}\sum\nolimits_{p}\sum\nolimits_{q}\mathrm{e}^{-2\pi\mathrm{i} (l p/N+ n q/T)}{\bar{u}}_{p,q}$, for $l=0, 1,..., N$ and $n=0, 1,..., T$ defining $T$ as the number of time steps.~This yields the numerical frequency-wavenumber spectrum $E^{\rm ST}(\kappa, \omega)$ \cite{Johnson1992}.\\
\indent \textit{Verification of the computational approach on the linear nondispersive problem}$-$For basic verification of the computational approach, we analysis a rod at the limits of $B\rightarrow0, r\rightarrow0$ to recover the linear dispersion relation $\omega=c\kappa$ from Eqs. (4) and (5). We set $r=0$ and choose a small amplitude, $B=0.0005$, instead of setting $B=0$ to avoid numerical instabilities in the simulations. The excitation profile considered is the same as the one studied in Fig. 3.~This condition generates a practically linear nondispersive wave. The space-time solution is shown in Fig. \ref{FigSM2a} and the corresponding energy spectrum is plotted in Fig. \ref{FigSM2b}; we clearly see that numerical energy spectrum perfectly coincides with the infinitesimal-strain dispersion relation, thus confirming the verification.\\

\bibliography{PRLBibTex}

\begin{thebibliography}{55}%
\makeatletter
\providecommand \@ifxundefined [1]{%
 \@ifx{#1\undefined}
}%
\providecommand \@ifnum [1]{%
 \ifnum #1\expandafter \@firstoftwo
 \else \expandafter \@secondoftwo
 \fi
}%
\providecommand \@ifx [1]{%
 \ifx #1\expandafter \@firstoftwo
 \else \expandafter \@secondoftwo
 \fi
}%
\providecommand \natexlab [1]{#1}%
\providecommand \enquote  [1]{``#1''}%
\providecommand \bibnamefont  [1]{#1}%
\providecommand \bibfnamefont [1]{#1}%
\providecommand \citenamefont [1]{#1}%
\providecommand \href@noop [0]{\@secondoftwo}%
\providecommand \href [0]{\begingroup \@sanitize@url \@href}%
\providecommand \@href[1]{\@@startlink{#1}\@@href}%
\providecommand \@@href[1]{\endgroup#1\@@endlink}%
\providecommand \@sanitize@url [0]{\catcode `\\12\catcode `\$12\catcode
  `\&12\catcode `\#12\catcode `\^12\catcode `\_12\catcode `\%12\relax}%
\providecommand \@@startlink[1]{}%
\providecommand \@@endlink[0]{}%
\providecommand \url  [0]{\begingroup\@sanitize@url \@url }%
\providecommand \@url [1]{\endgroup\@href {#1}{\urlprefix }}%
\providecommand \urlprefix  [0]{URL }%
\providecommand \Eprint [0]{\href }%
\providecommand \doibase [0]{http://dx.doi.org/}%
\providecommand \selectlanguage [0]{\@gobble}%
\providecommand \bibinfo  [0]{\@secondoftwo}%
\providecommand \bibfield  [0]{\@secondoftwo}%
\providecommand \translation [1]{[#1]}%
\providecommand \BibitemOpen [0]{}%
\providecommand \bibitemStop [0]{}%
\providecommand \bibitemNoStop [0]{.\EOS\space}%
\providecommand \EOS [0]{\spacefactor3000\relax}%
\providecommand \BibitemShut  [1]{\csname bibitem#1\endcsname}%
\let\auto@bib@innerbib\@empty
\bibitem [{\citenamefont {Whitham}(1974)}]{Whitham1974}%
  \BibitemOpen
  \bibfield  {author} {\bibinfo {author} {\bibfnamefont {G.~B.}\ \bibnamefont
  {Whitham}},\ }\href@noop {} {\emph {\bibinfo {title} {Linear and nonlinear
  waves}}}\ (\bibinfo  {publisher} {John Wiley \& Sons, New York},\ \bibinfo
  {year} {1974})\BibitemShut {NoStop}%
\bibitem [{\citenamefont {Nishijima}(1969)}]{Nishijima1969}%
  \BibitemOpen
  \bibfield  {author} {\bibinfo {author} {\bibfnamefont {K.}~\bibnamefont
  {Nishijima}},\ }\href@noop {} {\emph {\bibinfo {title} {Fields and Particles:
  Field Theory and Dispersion Relations}}}\ (\bibinfo  {publisher} {W. A.
  Benjamin, New York},\ \bibinfo {year} {1969})\BibitemShut {NoStop}%
\bibitem [{\citenamefont {Carretero-Gonz\'alez}\ \emph
  {et~al.}(2008)\citenamefont {Carretero-Gonz\'alez}, \citenamefont
  {Frantzeskakis},\ and\ \citenamefont {Kevrekidis}}]{Carretero2008}%
  \BibitemOpen
  \bibfield  {author} {\bibinfo {author} {\bibfnamefont {R.}~\bibnamefont
  {Carretero-Gonz\'alez}}, \bibinfo {author} {\bibfnamefont {D.}~\bibnamefont
  {Frantzeskakis}}, \ and\ \bibinfo {author} {\bibfnamefont {P.}~\bibnamefont
  {Kevrekidis}},\ }\href@noop {} {\bibfield  {journal} {\bibinfo  {journal}
  {Nonlinearity}\ }\textbf {\bibinfo {volume} {21}},\ \bibinfo {pages} {R139}
  (\bibinfo {year} {2008})}\BibitemShut {NoStop}%
\bibitem [{\citenamefont {Parker}(1985)}]{Parker1985}%
  \BibitemOpen
  \bibfield  {author} {\bibinfo {author} {\bibfnamefont {D.}~\bibnamefont
  {Parker}},\ }\href@noop {} {\bibfield  {journal} {\bibinfo  {journal}
  {Physica D}\ }\textbf {\bibinfo {volume} {16}},\ \bibinfo {pages} {358}
  (\bibinfo {year} {1985})}\BibitemShut {NoStop}%
\bibitem [{\citenamefont {Chakraborty}\ and\ \citenamefont
  {Mallik}(2001)}]{Chakraborty2001}%
  \BibitemOpen
  \bibfield  {author} {\bibinfo {author} {\bibfnamefont {G.}~\bibnamefont
  {Chakraborty}}\ and\ \bibinfo {author} {\bibfnamefont {A.}~\bibnamefont
  {Mallik}},\ }\href@noop {} {\bibfield  {journal} {\bibinfo  {journal} {Int.
  J. Nonlin. Mech.}\ }\textbf {\bibinfo {volume} {36}},\ \bibinfo {pages} {375}
  (\bibinfo {year} {2001})}\BibitemShut {NoStop}%
\bibitem [{\citenamefont {Cobelli}\ \emph {et~al.}(2009)\citenamefont
  {Cobelli}, \citenamefont {Petitjeans}, \citenamefont {Maurel}, \citenamefont
  {Pagneux},\ and\ \citenamefont {Mordant}}]{Cobelli2009}%
  \BibitemOpen
  \bibfield  {author} {\bibinfo {author} {\bibfnamefont {P.}~\bibnamefont
  {Cobelli}}, \bibinfo {author} {\bibfnamefont {P.}~\bibnamefont {Petitjeans}},
  \bibinfo {author} {\bibfnamefont {A.}~\bibnamefont {Maurel}}, \bibinfo
  {author} {\bibfnamefont {V.}~\bibnamefont {Pagneux}}, \ and\ \bibinfo
  {author} {\bibfnamefont {N.}~\bibnamefont {Mordant}},\ }\href@noop {}
  {\bibfield  {journal} {\bibinfo  {journal} {Phys. Rev. Lett.}\ }\textbf
  {\bibinfo {volume} {103}},\ \bibinfo {pages} {204301} (\bibinfo {year}
  {2009})}\BibitemShut {NoStop}%
\bibitem [{\citenamefont {Lee}\ \emph {et~al.}(2013)\citenamefont {Lee},
  \citenamefont {Kovačič},\ and\ \citenamefont {Cai}}]{Lee2013}%
  \BibitemOpen
  \bibfield  {author} {\bibinfo {author} {\bibfnamefont {W.}~\bibnamefont
  {Lee}}, \bibinfo {author} {\bibfnamefont {G.}~\bibnamefont {Kovačič}}, \
  and\ \bibinfo {author} {\bibfnamefont {D.}~\bibnamefont {Cai}},\ }\href@noop
  {} {\bibfield  {journal} {\bibinfo  {journal} {Proc. Natl. Acad. Sci.}\
  }\textbf {\bibinfo {volume} {110}},\ \bibinfo {pages} {3237} (\bibinfo {year}
  {2013})}\BibitemShut {NoStop}%
\bibitem [{\citenamefont {Shukla}\ \emph {et~al.}(2006)\citenamefont {Shukla},
  \citenamefont {Kourakis}, \citenamefont {Eliasson}, \citenamefont
  {Marklund},\ and\ \citenamefont {Stenflo}}]{Shukla2006}%
  \BibitemOpen
  \bibfield  {author} {\bibinfo {author} {\bibfnamefont {P.~K.}\ \bibnamefont
  {Shukla}}, \bibinfo {author} {\bibfnamefont {I.}~\bibnamefont {Kourakis}},
  \bibinfo {author} {\bibfnamefont {B.}~\bibnamefont {Eliasson}}, \bibinfo
  {author} {\bibfnamefont {M.}~\bibnamefont {Marklund}}, \ and\ \bibinfo
  {author} {\bibfnamefont {L.}~\bibnamefont {Stenflo}},\ }\href@noop {}
  {\bibfield  {journal} {\bibinfo  {journal} {Phys. Rev. Lett.}\ }\textbf
  {\bibinfo {volume} {97}},\ \bibinfo {pages} {094501} (\bibinfo {year}
  {2006})}\BibitemShut {NoStop}%
\bibitem [{\citenamefont {Onorato}\ \emph {et~al.}(2006)\citenamefont
  {Onorato}, \citenamefont {Osborne},\ and\ \citenamefont
  {Serio}}]{Onorato2006}%
  \BibitemOpen
  \bibfield  {author} {\bibinfo {author} {\bibfnamefont {M.}~\bibnamefont
  {Onorato}}, \bibinfo {author} {\bibfnamefont {A.}~\bibnamefont {Osborne}}, \
  and\ \bibinfo {author} {\bibfnamefont {M.}~\bibnamefont {Serio}},\
  }\href@noop {} {\bibfield  {journal} {\bibinfo  {journal} {Phys. Rev. Lett.}\
  }\textbf {\bibinfo {volume} {96}},\ \bibinfo {pages} {014503} (\bibinfo
  {year} {2006})}\BibitemShut {NoStop}%
\bibitem [{\citenamefont {Clark~di Leoni}\ \emph {et~al.}(2014)\citenamefont
  {Clark~di Leoni}, \citenamefont {Cobelli},\ and\ \citenamefont
  {Mininni}}]{Leoni2014}%
  \BibitemOpen
  \bibfield  {author} {\bibinfo {author} {\bibfnamefont {P.}~\bibnamefont
  {Clark~di Leoni}}, \bibinfo {author} {\bibfnamefont {P.}~\bibnamefont
  {Cobelli}}, \ and\ \bibinfo {author} {\bibfnamefont {P.}~\bibnamefont
  {Mininni}},\ }\href@noop {} {\bibfield  {journal} {\bibinfo  {journal} {Phys.
  Rev. E}\ }\textbf {\bibinfo {volume} {89}},\ \bibinfo {pages} {063025}
  (\bibinfo {year} {2014})}\BibitemShut {NoStop}%
\bibitem [{\citenamefont {Herbert}\ \emph {et~al.}(2010)\citenamefont
  {Herbert}, \citenamefont {Mordant},\ and\ \citenamefont
  {Falcon}}]{Herbert2010}%
  \BibitemOpen
  \bibfield  {author} {\bibinfo {author} {\bibfnamefont {E.}~\bibnamefont
  {Herbert}}, \bibinfo {author} {\bibfnamefont {N.}~\bibnamefont {Mordant}}, \
  and\ \bibinfo {author} {\bibfnamefont {E.}~\bibnamefont {Falcon}},\
  }\href@noop {} {\bibfield  {journal} {\bibinfo  {journal} {Phys. Rev. Lett.}\
  }\textbf {\bibinfo {volume} {105}},\ \bibinfo {pages} {144502} (\bibinfo
  {year} {2010})}\BibitemShut {NoStop}%
\bibitem [{\citenamefont {Gusev}\ \emph {et~al.}(1998)\citenamefont {Gusev},
  \citenamefont {Lauriks},\ and\ \citenamefont {Thoen}}]{Gusev1998}%
  \BibitemOpen
  \bibfield  {author} {\bibinfo {author} {\bibfnamefont {V.~E.}\ \bibnamefont
  {Gusev}}, \bibinfo {author} {\bibfnamefont {W.}~\bibnamefont {Lauriks}}, \
  and\ \bibinfo {author} {\bibfnamefont {J.}~\bibnamefont {Thoen}},\
  }\href@noop {} {\bibfield  {journal} {\bibinfo  {journal} {J. Acoust. Soc.
  Am.}\ }\textbf {\bibinfo {volume} {103}},\ \bibinfo {pages} {3216} (\bibinfo
  {year} {1998})}\BibitemShut {NoStop}%
\bibitem [{\citenamefont {Shadrivov}\ \emph {et~al.}(2004)\citenamefont
  {Shadrivov}, \citenamefont {Sukhorukov}, \citenamefont {Kivshar},
  \citenamefont {Zharov}, \citenamefont {Boardman},\ and\ \citenamefont
  {Egan}}]{Shadrivov2004}%
  \BibitemOpen
  \bibfield  {author} {\bibinfo {author} {\bibfnamefont {I.~V.}\ \bibnamefont
  {Shadrivov}}, \bibinfo {author} {\bibfnamefont {A.~A.}\ \bibnamefont
  {Sukhorukov}}, \bibinfo {author} {\bibfnamefont {Y.~S.}\ \bibnamefont
  {Kivshar}}, \bibinfo {author} {\bibfnamefont {A.~A.}\ \bibnamefont {Zharov}},
  \bibinfo {author} {\bibfnamefont {A.~D.}\ \bibnamefont {Boardman}}, \ and\
  \bibinfo {author} {\bibfnamefont {P.}~\bibnamefont {Egan}},\ }\href@noop {}
  {\bibfield  {journal} {\bibinfo  {journal} {Phys. Rev. E}\ }\textbf {\bibinfo
  {volume} {69}},\ \bibinfo {pages} {016617} (\bibinfo {year}
  {2004})}\BibitemShut {NoStop}%
\bibitem [{\citenamefont {Kourakis}\ and\ \citenamefont
  {Shukla}(2005)}]{Kourakis2005}%
  \BibitemOpen
  \bibfield  {author} {\bibinfo {author} {\bibfnamefont {I.}~\bibnamefont
  {Kourakis}}\ and\ \bibinfo {author} {\bibfnamefont {P.~K.}\ \bibnamefont
  {Shukla}},\ }\href@noop {} {\bibfield  {journal} {\bibinfo  {journal} {Phys.
  Rev. E}\ }\textbf {\bibinfo {volume} {72}},\ \bibinfo {pages} {016626}
  (\bibinfo {year} {2005})}\BibitemShut {NoStop}%
\bibitem [{\citenamefont {Yoon}\ \emph {et~al.}(2003)\citenamefont {Yoon},
  \citenamefont {Gaelzer}, \citenamefont {Umeda}, \citenamefont {Omura},\ and\
  \citenamefont {Matsumoto}}]{Yoon2003}%
  \BibitemOpen
  \bibfield  {author} {\bibinfo {author} {\bibfnamefont {P.}~\bibnamefont
  {Yoon}}, \bibinfo {author} {\bibfnamefont {R.}~\bibnamefont {Gaelzer}},
  \bibinfo {author} {\bibfnamefont {T.}~\bibnamefont {Umeda}}, \bibinfo
  {author} {\bibfnamefont {Y.}~\bibnamefont {Omura}}, \ and\ \bibinfo {author}
  {\bibfnamefont {H.}~\bibnamefont {Matsumoto}},\ }\href@noop {} {\bibfield
  {journal} {\bibinfo  {journal} {Phys. Plasmas}\ }\textbf {\bibinfo {volume}
  {10}},\ \bibinfo {pages} {364} (\bibinfo {year} {2003})}\BibitemShut
  {NoStop}%
\bibitem [{\citenamefont {Huang}\ \emph {et~al.}(2009)\citenamefont {Huang},
  \citenamefont {Chang}, \citenamefont {Leung},\ and\ \citenamefont
  {Tsai}}]{Huang2009}%
  \BibitemOpen
  \bibfield  {author} {\bibinfo {author} {\bibfnamefont {J.-H.}\ \bibnamefont
  {Huang}}, \bibinfo {author} {\bibfnamefont {R.}~\bibnamefont {Chang}},
  \bibinfo {author} {\bibfnamefont {P.-T.}\ \bibnamefont {Leung}}, \ and\
  \bibinfo {author} {\bibfnamefont {D.~P.}\ \bibnamefont {Tsai}},\ }\href@noop
  {} {\bibfield  {journal} {\bibinfo  {journal} {Opt. Commun.}\ }\textbf
  {\bibinfo {volume} {282}},\ \bibinfo {pages} {1412} (\bibinfo {year}
  {2009})}\BibitemShut {NoStop}%
\bibitem [{\citenamefont {Ginzburg}\ \emph {et~al.}(2010)\citenamefont
  {Ginzburg}, \citenamefont {Hayat}, \citenamefont {Berkovitch},\ and\
  \citenamefont {Orenstein}}]{Ginzburg2010}%
  \BibitemOpen
  \bibfield  {author} {\bibinfo {author} {\bibfnamefont {P.}~\bibnamefont
  {Ginzburg}}, \bibinfo {author} {\bibfnamefont {A.}~\bibnamefont {Hayat}},
  \bibinfo {author} {\bibfnamefont {N.}~\bibnamefont {Berkovitch}}, \ and\
  \bibinfo {author} {\bibfnamefont {M.}~\bibnamefont {Orenstein}},\ }\href@noop
  {} {\bibfield  {journal} {\bibinfo  {journal} {Opt. Lett.}\ }\textbf
  {\bibinfo {volume} {35}},\ \bibinfo {pages} {1551} (\bibinfo {year}
  {2010})}\BibitemShut {NoStop}%
\bibitem [{\citenamefont {Hager}\ and\ \citenamefont
  {Hallatschek}(2012)}]{Hager2012}%
  \BibitemOpen
  \bibfield  {author} {\bibinfo {author} {\bibfnamefont {R.}~\bibnamefont
  {Hager}}\ and\ \bibinfo {author} {\bibfnamefont {K.}~\bibnamefont
  {Hallatschek}},\ }\href@noop {} {\bibfield  {journal} {\bibinfo  {journal}
  {Phys. Rev. Lett.}\ }\textbf {\bibinfo {volume} {108}},\ \bibinfo {pages}
  {035004} (\bibinfo {year} {2012})}\BibitemShut {NoStop}%
\bibitem [{\citenamefont {Fritts}\ and\ \citenamefont
  {Alexander}(2003)}]{Fritts2003}%
  \BibitemOpen
  \bibfield  {author} {\bibinfo {author} {\bibfnamefont {D.}~\bibnamefont
  {Fritts}}\ and\ \bibinfo {author} {\bibfnamefont {M.}~\bibnamefont
  {Alexander}},\ }\href@noop {} {\bibfield  {journal} {\bibinfo  {journal}
  {Rev. Geophys.}\ }\textbf {\bibinfo {volume} {41}},\ \bibinfo {pages} {1003}
  (\bibinfo {year} {2003})}\BibitemShut {NoStop}%
\bibitem [{\citenamefont {Debnath}(2007)}]{Debnath2007}%
  \BibitemOpen
  \bibfield  {author} {\bibinfo {author} {\bibfnamefont {L.}~\bibnamefont
  {Debnath}},\ }\href@noop {} {\bibfield  {journal} {\bibinfo  {journal} {J.
  Math. Anal. Appl.}\ }\textbf {\bibinfo {volume} {333}},\ \bibinfo {pages}
  {164–190} (\bibinfo {year} {2007})}\BibitemShut {NoStop}%
\bibitem [{\citenamefont {Boyd}(2018)}]{Boyd2018}%
  \BibitemOpen
  \bibfield  {author} {\bibinfo {author} {\bibfnamefont {J.}~\bibnamefont
  {Boyd}},\ }in\ \href@noop {} {\emph {\bibinfo {booktitle} {Dynamics of the
  Equatorial Ocean}}},\ \bibinfo {editor} {edited by\ \bibinfo {editor}
  {\bibfnamefont {J.}~\bibnamefont {Boyd}}}\ (\bibinfo  {publisher}
  {Springer},\ \bibinfo {address} {Berlin},\ \bibinfo {year} {2018})\ pp.\
  \bibinfo {pages} {329--404}\BibitemShut {NoStop}%
\bibitem [{\citenamefont {Davydov}(2018)}]{Davydov1973}%
  \BibitemOpen
  \bibfield  {author} {\bibinfo {author} {\bibfnamefont {A.}~\bibnamefont
  {Davydov}},\ }\href@noop {} {\bibfield  {journal} {\bibinfo  {journal} {J.
  Theor. Biol}\ }\textbf {\bibinfo {volume} {38}},\ \bibinfo {pages} {559}
  (\bibinfo {year} {2018})}\BibitemShut {NoStop}%
\bibitem [{\citenamefont {Mvogo}\ \emph {et~al.}(2018)\citenamefont {Mvogo},
  \citenamefont {Ben-Bolie},\ and\ \citenamefont {Kofané}}]{Mvogo2018}%
  \BibitemOpen
  \bibfield  {author} {\bibinfo {author} {\bibfnamefont {A.}~\bibnamefont
  {Mvogo}}, \bibinfo {author} {\bibfnamefont {G.}~\bibnamefont {Ben-Bolie}}, \
  and\ \bibinfo {author} {\bibfnamefont {T.}~\bibnamefont {Kofané}},\
  }\href@noop {} {\bibfield  {journal} {\bibinfo  {journal} {Eur. Phys. J. B}\
  }\textbf {\bibinfo {volume} {86}},\ \bibinfo {pages} {217} (\bibinfo {year}
  {2018})}\BibitemShut {NoStop}%
\bibitem [{\citenamefont {Gardner}\ \emph {et~al.}(1967)\citenamefont
  {Gardner}, \citenamefont {Greene},\ and\ \citenamefont
  {Miura}}]{Gardner1967}%
  \BibitemOpen
  \bibfield  {author} {\bibinfo {author} {\bibfnamefont {C.}~\bibnamefont
  {Gardner}}, \bibinfo {author} {\bibfnamefont {K.~M.}\ \bibnamefont {Greene},
  \bibfnamefont {J.M.}}, \ and\ \bibinfo {author} {\bibfnamefont
  {R.}~\bibnamefont {Miura}},\ }\href@noop {} {\bibfield  {journal} {\bibinfo
  {journal} {Phys. Rev. Lett.}\ }\textbf {\bibinfo {volume} {19}},\ \bibinfo
  {pages} {1095} (\bibinfo {year} {1967})}\BibitemShut {NoStop}%
\bibitem [{\citenamefont {Lax}(1968)}]{Lax1968}%
  \BibitemOpen
  \bibfield  {author} {\bibinfo {author} {\bibfnamefont {P.}~\bibnamefont
  {Lax}},\ }\href@noop {} {\bibfield  {journal} {\bibinfo  {journal} {Commun.
  Pur. Appl. Math.}\ }\textbf {\bibinfo {volume} {XXI}},\ \bibinfo {pages}
  {467} (\bibinfo {year} {1968})}\BibitemShut {NoStop}%
\bibitem [{\citenamefont {Zakharov}\ and\ \citenamefont
  {Faddeev}(1971)}]{Zakharov1971}%
  \BibitemOpen
  \bibfield  {author} {\bibinfo {author} {\bibfnamefont {V.}~\bibnamefont
  {Zakharov}}\ and\ \bibinfo {author} {\bibfnamefont {L.}~\bibnamefont
  {Faddeev}},\ }\href@noop {} {\bibfield  {journal} {\bibinfo  {journal}
  {Funct. Anal. Appl.}\ }\textbf {\bibinfo {volume} {5}},\ \bibinfo {pages}
  {280} (\bibinfo {year} {1971})}\BibitemShut {NoStop}%
\bibitem [{\citenamefont {Ablowitz}\ \emph {et~al.}(1974)\citenamefont
  {Ablowitz}, \citenamefont {Kaup}, \citenamefont {Newell},\ and\ \citenamefont
  {Segur}}]{Ablowitz1974}%
  \BibitemOpen
  \bibfield  {author} {\bibinfo {author} {\bibfnamefont {M.}~\bibnamefont
  {Ablowitz}}, \bibinfo {author} {\bibfnamefont {D.}~\bibnamefont {Kaup}},
  \bibinfo {author} {\bibfnamefont {A.}~\bibnamefont {Newell}}, \ and\ \bibinfo
  {author} {\bibfnamefont {H.}~\bibnamefont {Segur}},\ }\href@noop {}
  {\bibfield  {journal} {\bibinfo  {journal} {Stud. Appl. Math.}\ }\textbf
  {\bibinfo {volume} {53}},\ \bibinfo {pages} {249} (\bibinfo {year}
  {1974})}\BibitemShut {NoStop}%
\bibitem [{\citenamefont {Whitham}(1965)}]{Whitham1965}%
  \BibitemOpen
  \bibfield  {author} {\bibinfo {author} {\bibfnamefont {G.~B.}\ \bibnamefont
  {Whitham}},\ }\href@noop {} {\bibfield  {journal} {\bibinfo  {journal} {Proc.
  R. Soc. A}\ }\textbf {\bibinfo {volume} {283}},\ \bibinfo {pages} {238}
  (\bibinfo {year} {1965})}\BibitemShut {NoStop}%
\bibitem [{\citenamefont {Sch\"urmann}\ \emph {et~al.}(1998)\citenamefont
  {Sch\"urmann}, \citenamefont {Serov},\ and\ \citenamefont
  {Shestopalov}}]{Schurmann1988}%
  \BibitemOpen
  \bibfield  {author} {\bibinfo {author} {\bibfnamefont {H.}~\bibnamefont
  {Sch\"urmann}}, \bibinfo {author} {\bibfnamefont {V.}~\bibnamefont {Serov}},
  \ and\ \bibinfo {author} {\bibfnamefont {Y.}~\bibnamefont {Shestopalov}},\
  }\href@noop {} {\bibfield  {journal} {\bibinfo  {journal} {Phys. Rev. E}\
  }\textbf {\bibinfo {volume} {58}},\ \bibinfo {pages} {1040} (\bibinfo {year}
  {1998})}\BibitemShut {NoStop}%
\bibitem [{\citenamefont {Franken}\ \emph {et~al.}(1961)\citenamefont
  {Franken}, \citenamefont {Hill}, \citenamefont {Peters},\ and\ \citenamefont
  {Weinreich}}]{Franken1961}%
  \BibitemOpen
  \bibfield  {author} {\bibinfo {author} {\bibfnamefont {P.}~\bibnamefont
  {Franken}}, \bibinfo {author} {\bibfnamefont {A.}~\bibnamefont {Hill}},
  \bibinfo {author} {\bibfnamefont {C.}~\bibnamefont {Peters}}, \ and\ \bibinfo
  {author} {\bibfnamefont {G.}~\bibnamefont {Weinreich}},\ }\href@noop {}
  {\bibfield  {journal} {\bibinfo  {journal} {Phys. Rev. Lett.}\ }\textbf
  {\bibinfo {volume} {7}},\ \bibinfo {pages} {118} (\bibinfo {year}
  {1961})}\BibitemShut {NoStop}%
\bibitem [{\citenamefont {Hasselmann}(1962)}]{Hasselmann1962}%
  \BibitemOpen
  \bibfield  {author} {\bibinfo {author} {\bibfnamefont {K.}~\bibnamefont
  {Hasselmann}},\ }\href@noop {} {\bibfield  {journal} {\bibinfo  {journal} {J.
  Fluid Mech.}\ }\textbf {\bibinfo {volume} {12}},\ \bibinfo {pages} {481}
  (\bibinfo {year} {1962})}\BibitemShut {NoStop}%
\bibitem [{\citenamefont {Benney}\ and\ \citenamefont
  {Saffman}(1966)}]{Benney1996}%
  \BibitemOpen
  \bibfield  {author} {\bibinfo {author} {\bibfnamefont {D.}~\bibnamefont
  {Benney}}\ and\ \bibinfo {author} {\bibfnamefont {P.}~\bibnamefont
  {Saffman}},\ }\href@noop {} {\bibfield  {journal} {\bibinfo  {journal} {Proc.
  R. Soc. A}\ }\textbf {\bibinfo {volume} {289}},\ \bibinfo {pages} {301}
  (\bibinfo {year} {1966})}\BibitemShut {NoStop}%
\bibitem [{\citenamefont {Thurston}\ and\ \citenamefont
  {Shapiro}(1967)}]{Thursten1967}%
  \BibitemOpen
  \bibfield  {author} {\bibinfo {author} {\bibfnamefont {R.}~\bibnamefont
  {Thurston}}\ and\ \bibinfo {author} {\bibfnamefont {M.}~\bibnamefont
  {Shapiro}},\ }\href@noop {} {\bibfield  {journal} {\bibinfo  {journal} {J.
  Acoust. Soc. Am.}\ }\textbf {\bibinfo {volume} {41}},\ \bibinfo {pages}
  {1112} (\bibinfo {year} {1967})}\BibitemShut {NoStop}%
\bibitem [{\citenamefont {Tiersten}\ and\ \citenamefont
  {Baumhauer}(1974)}]{Tiersten1974}%
  \BibitemOpen
  \bibfield  {author} {\bibinfo {author} {\bibfnamefont {H.}~\bibnamefont
  {Tiersten}}\ and\ \bibinfo {author} {\bibfnamefont {J.}~\bibnamefont
  {Baumhauer}},\ }\href@noop {} {\bibfield  {journal} {\bibinfo  {journal} {J.
  Appl. Phys.}\ }\textbf {\bibinfo {volume} {45}},\ \bibinfo {pages} {4272}
  (\bibinfo {year} {1974})}\BibitemShut {NoStop}%
\bibitem [{\citenamefont {Thompson}\ and\ \citenamefont
  {Tiersten}(1977)}]{Thompson1977}%
  \BibitemOpen
  \bibfield  {author} {\bibinfo {author} {\bibfnamefont {R.}~\bibnamefont
  {Thompson}}\ and\ \bibinfo {author} {\bibfnamefont {H.}~\bibnamefont
  {Tiersten}},\ }\href@noop {} {\bibfield  {journal} {\bibinfo  {journal} {J.
  Acoust. Soc. Am.}\ }\textbf {\bibinfo {volume} {62}},\ \bibinfo {pages} {33}
  (\bibinfo {year} {1977})}\BibitemShut {NoStop}%
\bibitem [{\citenamefont {Auld}(1973)}]{Auld1973}%
  \BibitemOpen
  \bibfield  {author} {\bibinfo {author} {\bibfnamefont {B.}~\bibnamefont
  {Auld}},\ }\href@noop {} {\emph {\bibinfo {title} {Acoustic Fields and Waves
  in Solids, Vols. I and II}}}\ (\bibinfo  {publisher} {Wiley, London},\
  \bibinfo {year} {1973})\BibitemShut {NoStop}%
\bibitem [{\citenamefont {Deng}(1999)}]{Deng1999}%
  \BibitemOpen
  \bibfield  {author} {\bibinfo {author} {\bibfnamefont {M.}~\bibnamefont
  {Deng}},\ }\href@noop {} {\bibfield  {journal} {\bibinfo  {journal} {J. Appl.
  Phys.}\ }\textbf {\bibinfo {volume} {85}},\ \bibinfo {pages} {3051} (\bibinfo
  {year} {1999})}\BibitemShut {NoStop}%
\bibitem [{\citenamefont {de~Lima}\ and\ \citenamefont
  {Hamilton}(2003)}]{Lima2003}%
  \BibitemOpen
  \bibfield  {author} {\bibinfo {author} {\bibfnamefont {W.}~\bibnamefont
  {de~Lima}}\ and\ \bibinfo {author} {\bibfnamefont {M.}~\bibnamefont
  {Hamilton}},\ }\href@noop {} {\bibfield  {journal} {\bibinfo  {journal} {J.
  Sound Vib.}\ }\textbf {\bibinfo {volume} {265}},\ \bibinfo {pages}
  {819–839} (\bibinfo {year} {2003})}\BibitemShut {NoStop}%
\bibitem [{\citenamefont {Breazeale}\ and\ \citenamefont
  {Thompson}(1963)}]{Breazeale1963}%
  \BibitemOpen
  \bibfield  {author} {\bibinfo {author} {\bibfnamefont {M.}~\bibnamefont
  {Breazeale}}\ and\ \bibinfo {author} {\bibfnamefont {D.}~\bibnamefont
  {Thompson}},\ }\href@noop {} {\bibfield  {journal} {\bibinfo  {journal}
  {Appl. Phys. Lett.}\ }\textbf {\bibinfo {volume} {3}},\ \bibinfo {pages} {77}
  (\bibinfo {year} {1963})}\BibitemShut {NoStop}%
\bibitem [{\citenamefont {Hikata}\ \emph {et~al.}(1965)\citenamefont {Hikata},
  \citenamefont {Chick},\ and\ \citenamefont {Elbaum}}]{Hikata1965}%
  \BibitemOpen
  \bibfield  {author} {\bibinfo {author} {\bibfnamefont {A.}~\bibnamefont
  {Hikata}}, \bibinfo {author} {\bibfnamefont {B.}~\bibnamefont {Chick}}, \
  and\ \bibinfo {author} {\bibfnamefont {C.}~\bibnamefont {Elbaum}},\
  }\href@noop {} {\bibfield  {journal} {\bibinfo  {journal} {J. Appl. Phys.}\
  }\textbf {\bibinfo {volume} {36}},\ \bibinfo {pages} {229} (\bibinfo {year}
  {1965})}\BibitemShut {NoStop}%
\bibitem [{\citenamefont {Abedinnasab}\ and\ \citenamefont
  {Hussein}(2013)}]{Abedinnasab2013}%
  \BibitemOpen
  \bibfield  {author} {\bibinfo {author} {\bibfnamefont {M.~H.}\ \bibnamefont
  {Abedinnasab}}\ and\ \bibinfo {author} {\bibfnamefont {M.~I.}\ \bibnamefont
  {Hussein}},\ }\href@noop {} {\bibfield  {journal} {\bibinfo  {journal} {Wave
  Motion}\ }\textbf {\bibinfo {volume} {50}},\ \bibinfo {pages} {374} (\bibinfo
  {year} {2013})}\BibitemShut {NoStop}%
\bibitem [{\citenamefont {Seth}(1962)}]{Seth1962}%
  \BibitemOpen
  \bibfield  {author} {\bibinfo {author} {\bibfnamefont {B.~R.}\ \bibnamefont
  {Seth}},\ }\href@noop {} {\bibfield  {journal} {\bibinfo  {journal} {IUTAM
  Symposium on Second Order Effects in Elasticity, Plasticity and Fluid
  Mechanics, Haifa}\ ,\ \bibinfo {pages} {1}} (\bibinfo {year}
  {1962})}\BibitemShut {NoStop}%
\bibitem [{\citenamefont {Hill}(1968)}]{Hill1968}%
  \BibitemOpen
  \bibfield  {author} {\bibinfo {author} {\bibfnamefont {R.}~\bibnamefont
  {Hill}},\ }\href@noop {} {\bibfield  {journal} {\bibinfo  {journal} {Journal
  of the Mechanics and Physics of Solids}\ }\textbf {\bibinfo {volume} {16}},\
  \bibinfo {pages} {229} (\bibinfo {year} {1968})}\BibitemShut {NoStop}%
\bibitem [{Note1()}]{Note1}%
  \BibitemOpen
  \bibinfo {note} {For simplicity, we consider the effect of the lateral
  inertia only on the longitudinal displacement; however the theoretical
  framework is fully valid in the absence of this modeling
  simplification.}\BibitemShut {Stop}%
\bibitem [{Note2()}]{Note2}%
  \BibitemOpen
  \bibinfo {note} {All reported units are in the SI system.}\BibitemShut
  {Stop}%
\bibitem [{Note3()}]{Note3}%
  \BibitemOpen
  \bibinfo {note} {All displayed numerical contour plots are of the quantity
  $\protect \mathrm {ln}{ |E^{\protect \rm ST}_{B_{\protect \rm e},\kappa
  _{\protect \rm e}}|}$ (energy spectrum due to an excitation at $B_{\protect
  \rm e}$ and $\kappa _{\protect \rm e}$) or $\DOTSB \sum@ \slimits@ _{\kappa
  _{\protect \rm e}}[\protect \mathrm {ln}{ |E^{\protect \rm ST}_{B_{\protect
  \rm e},\kappa _{\protect \rm e}}|}]$ (superposition of energy spectra over
  several values of $\kappa _{\protect \rm e}$ for a given value of
  $B_{\protect \rm e}$). The Fourier transformation is always done at a time
  close to $\tau _\protect \mathrm {B}$. Details on the derivation of
  $E^{\protect \rm ST}$ are provided in Appendix A.}\BibitemShut {Stop}%
\bibitem [{\citenamefont {Ablowitz}(2011)}]{ablowitz2011nonlinear}%
  \BibitemOpen
  \bibfield  {author} {\bibinfo {author} {\bibfnamefont {M.~J.}\ \bibnamefont
  {Ablowitz}},\ }\href@noop {} {\emph {\bibinfo {title} {Nonlinear dispersive
  waves: Asymptotic analysis and solitons}}}\ (\bibinfo  {publisher} {Cambridge
  University Press, Cambridge},\ \bibinfo {year} {2011})\BibitemShut {NoStop}%
\bibitem [{Note4()}]{Note4}%
  \BibitemOpen
  \bibinfo {note} {The spatial profile of the waves feature the eventual
  formation of shocks at the leading and trailing edges of the wave packet for
  the GLS and HS measures, respectively$-$in analogy with the behavior observed
  in Fig.~\ref {Fig1a}.}\BibitemShut {Stop}%
\bibitem [{\citenamefont {Russell}(1844)}]{Russel1844}%
  \BibitemOpen
  \bibfield  {author} {\bibinfo {author} {\bibfnamefont {J.~S.}\ \bibnamefont
  {Russell}},\ }in\ \href@noop {} {\emph {\bibinfo {booktitle} {14th Meeting of
  the British Association for the Advancement of Science, York}}}\ (\bibinfo
  {year} {1844})\ pp.\ \bibinfo {pages} {311--390}\BibitemShut {NoStop}%
\bibitem [{\citenamefont {Korteweg}\ and\ \citenamefont
  {de~Vries}()}]{Korteweg1895}%
  \BibitemOpen
  \bibfield  {author} {\bibinfo {author} {\bibfnamefont {D.}~\bibnamefont
  {Korteweg}}\ and\ \bibinfo {author} {\bibfnamefont {G.}~\bibnamefont
  {de~Vries}},\ }\href@noop {} {\bibfield  {journal} {\bibinfo  {journal}
  {Phil. Mag.}\ }\textbf {\bibinfo {volume} {39}}}\BibitemShut {NoStop}%
\bibitem [{\citenamefont {Adlam}\ and\ \citenamefont {Allen}()}]{Adlam1958}%
  \BibitemOpen
  \bibfield  {author} {\bibinfo {author} {\bibfnamefont {J.~H.}\ \bibnamefont
  {Adlam}}\ and\ \bibinfo {author} {\bibfnamefont {J.~E.}\ \bibnamefont
  {Allen}},\ }\href@noop {} {\bibfield  {journal} {\bibinfo  {journal} {Phil.
  Mag.}\ }\textbf {\bibinfo {volume} {3}}}\BibitemShut {NoStop}%
\bibitem [{\citenamefont {Hussein}\ and\ \citenamefont
  {Khajehtourian}(2018)}]{Hussein2018}%
  \BibitemOpen
  \bibfield  {author} {\bibinfo {author} {\bibfnamefont {M.~I.}\ \bibnamefont
  {Hussein}}\ and\ \bibinfo {author} {\bibfnamefont {R.}~\bibnamefont
  {Khajehtourian}},\ }\href@noop {} {\bibfield  {journal} {\bibinfo  {journal}
  {Proc. R. Soc. A}\ }\textbf {\bibinfo {volume} {474}},\ \bibinfo {pages}
  {20180173} (\bibinfo {year} {2018})}\BibitemShut {NoStop}%
\bibitem [{\citenamefont {Remillieux}\ \emph {et~al.}(2016)\citenamefont
  {Remillieux}, \citenamefont {Guyer}, \citenamefont {Payan},\ and\
  \citenamefont {Ulrich}}]{Remillieux2016}%
  \BibitemOpen
  \bibfield  {author} {\bibinfo {author} {\bibfnamefont {M.~C.}\ \bibnamefont
  {Remillieux}}, \bibinfo {author} {\bibfnamefont {R.~A.}\ \bibnamefont
  {Guyer}}, \bibinfo {author} {\bibfnamefont {C.}~\bibnamefont {Payan}}, \ and\
  \bibinfo {author} {\bibfnamefont {T.}~\bibnamefont {Ulrich}},\ }\href@noop {}
  {\bibfield  {journal} {\bibinfo  {journal} {Phys. Rev. Lett.}\ }\textbf
  {\bibinfo {volume} {116}},\ \bibinfo {pages} {115501} (\bibinfo {year}
  {2016})}\BibitemShut {NoStop}%
\bibitem [{\citenamefont {Dai}\ and\ \citenamefont {Huo}(2000)}]{Dai2000}%
  \BibitemOpen
  \bibfield  {author} {\bibinfo {author} {\bibfnamefont {H.-H.}\ \bibnamefont
  {Dai}}\ and\ \bibinfo {author} {\bibfnamefont {Y.}~\bibnamefont {Huo}},\
  }\href@noop {} {\bibfield  {journal} {\bibinfo  {journal} {Proc. R. Soc. A}\
  }\textbf {\bibinfo {volume} {456}},\ \bibinfo {pages} {331–} (\bibinfo
  {year} {2000})}\BibitemShut {NoStop}%
\bibitem [{\citenamefont {Johnson}\ and\ \citenamefont
  {Dudgeon}(1992)}]{Johnson1992}%
  \BibitemOpen
  \bibfield  {author} {\bibinfo {author} {\bibfnamefont {D.~H.}\ \bibnamefont
  {Johnson}}\ and\ \bibinfo {author} {\bibfnamefont {D.~E.}\ \bibnamefont
  {Dudgeon}},\ }\href@noop {} {\emph {\bibinfo {title} {Array signal
  processing: concepts and techniques}}}\ (\bibinfo  {publisher} {Simon \&
  Schuster},\ \bibinfo {year} {1992})\BibitemShut {NoStop}%
\end{thebibliography}%
\end{document}